\documentclass[aps,prx, superscriptaddress,twocolumn]{revtex4-1}
\usepackage[utf8]{inputenc}
\usepackage[T1]{fontenc}
\usepackage[francais, english]{babel}
\usepackage{amssymb}
\usepackage{amsmath,amsfonts,color,bm}
\usepackage{esvect}
\usepackage{graphicx}
\usepackage{fancyhdr}
\usepackage{makeidx}
\usepackage{pst-tree}
\usepackage{epstopdf}
\usepackage{vmargin}  
\usepackage{latexsym}
\usepackage{graphicx}
\usepackage{amssymb}
\usepackage{epstopdf}
\usepackage{wrapfig}
\usepackage{caption}
\usepackage{gensymb}

\DeclareCaptionLabelFormat{suppl}{#1 S#2}%
\newcommand{\NK}[1]{{\color{black} {#1}}}
\newcommand{\ML}[1]{{\color{black} {#1}}}

\newcommand{\MLcorr}[1]{{\color{black} {#1}}}
\newcommand{\REF}[1]{{\color{black} {#1}}}
\newcommand{\CORR}[1]{{\color{black} {#1}}}

\begin{document}

\title{Phonon drag-electric current generation at the liquid-graphite interface}
\title{Rising electronic winds under liquid flows on carbon surfaces}
\title{Electronic winds risen under liquid flows on carbon surfaces}
\title{Strong electronic winds blowing under liquid flows on carbon surfaces}

\author{Mathieu Liz\'ee}
\thanks{These authors contributed equally to this work}
\address{Laboratoire de Physique de l'\'Ecole Normale Sup\'erieure, ENS, Universit\'e PSL, CNRS, 3 Sorbonne Universit\'e, Universit\'e Paris Cit\'e, 75005 Paris, France}
\author{Alice Marcotte} 
\thanks{These authors contributed equally to this work}
\address{Laboratoire de Physique de l'\'Ecole Normale Sup\'erieure, ENS, Universit\'e PSL, CNRS, 3 Sorbonne Universit\'e, Universit\'e Paris Cit\'e, 75005 Paris, France}
\author{Baptiste Coquinot}
\address{Laboratoire de Physique de l'\'Ecole Normale Sup\'erieure, ENS, Universit\'e PSL, CNRS, 3 Sorbonne Universit\'e, Universit\'e Paris Cit\'e, 75005 Paris, France}
\address{Center for Computational Quantum Physics, Flatiron Institute, New York,  USA}
\author{Nikita Kavokine}
\address{Center for Computational Quantum Physics, Flatiron Institute, New York,  USA}
\author{Karen Sobnath}
\address{Universit\'e Paris Cit\'e, Laboratoire Mat\'eriaux et Ph\'enomènes Quantiques, CNRS, F-75013 Paris, France}
\author{Cl\'ement Barraud}
\address{Universit\'e Paris Cit\'e, Laboratoire Mat\'eriaux et Ph\'enomènes Quantiques, CNRS, F-75013 Paris, France}
\author{Ankit Bhardwaj}
\address{National Graphene Institute, The University of Manchester, UK}
\address{Department of Physics and Astronomy, The University of Manchester, UK}
\author{Boya Radha}
\address{National Graphene Institute, The University of Manchester, UK}
\address{Department of Physics and Astronomy, The University of Manchester, UK}
\author{Antoine Nigu\`es}
\address{Laboratoire de Physique de l'\'Ecole Normale Sup\'erieure, ENS, Universit\'e PSL, CNRS, 3 Sorbonne Universit\'e, Universit\'e Paris Cit\'e, 75005 Paris, France}
\author{Lyd\'eric Bocquet} 
\address{Laboratoire de Physique de l'\'Ecole Normale Sup\'erieure, ENS, Universit\'e PSL, CNRS, 3 Sorbonne Universit\'e, Universit\'e Paris Cit\'e, 75005 Paris, France}
\author{Alessandro Siria}
\address{Laboratoire de Physique de l'\'Ecole Normale Sup\'erieure, ENS, Universit\'e PSL, CNRS, 3 Sorbonne Universit\'e, Universit\'e Paris Cit\'e, 75005 Paris, France}


\begin{abstract}

\NK{Solid-liquid interfaces display a wealth of emerging phenomena at nanometer scales, which are at the root of their technological applications. While the interfacial structure and chemistry have been intensively explored, the potential coupling between liquid flows and the solid's electronic degrees of freedom has been broadly overlooked up till now. Despite several reports of electronic currents induced by liquids flowing in various carbon nanostructures, the mechanisms at stake remain controversial. Here, we unveil the molecular mechanisms of interfacial liquid-electron coupling by investigating flow-induced current generation at the nanoscale. We use a tuning fork Atomic Force Microscope (AFM) to deposit and displace a micrometric liquid droplet on a multilayer graphene sample, and observe an electronic current induced by the droplet displacement. The measured current is several orders of magnitude larger than previously reported for water on carbon, and further boosted by the presence of surface wrinkles on the carbon surface. Our results point to a peculiar momentum transfer mechanism between the fluid molecules and graphene charge carriers, mediated mainly by the solid's phonon excitations. These findings open new avenues for active control of nanoscale liquid flows through the solid walls' electronic degrees of freedom. }

\end{abstract}

\maketitle

\section{Introduction}

The transport of liquids near carbon surfaces has unveiled a wide range of unexpected properties, ranging from very fast permeation to non-linear ion transport and sieving \cite{Kavokine2021,Faucher2019},  \ML{as well as promising applications in energy harvesting \cite{Zhang2018}.} These results highlight the peculiar nature of the liquid-solid interaction. \NK{In a classical framework -- such as the widely used force-field molecular dynamics simulations -- a solid acts on fluid molecules as a static periodic potential, and the interfacial fluid dynamics are understood in terms of the atomic-scale surface roughness. Yet, a number of recent observations have challenged this conventional description by revealing couplings between liquid flows and the solid's electronic degrees of freedom. Prominent examples include liquid-flow-induced electronic currents  \cite{Gosh2003, Cohen2003, Zhao2008, Rabinowitz2020,Dhiman2011, Yin2012, Lee2013,Kuriya2020, Yin2014, Park2017}}, the modification of wetting by electronic screening \cite{Comtet2017,Schlaich2022}, or anomalies in the water-carbon friction \cite{Kavokine2022}. \NK{This points to the need of understanding the molecular mechanisms at stake: in particular, several observations of flow-induced electronic currents \cite{Rabinowitz2020} still lack a satisfactory rationalization.} 

\NK{In this article, we report the generation of an electric current in a few-layer graphene sample by the displacement of a liquid droplet. As opposed to previous reports, our experiment has sub-micrometer dimensions: such dimensions are incompatible with known current generation mechanisms, such as the charging/discharging of a pseudo-capacitance~\cite{Yin2014,Park2017}. With the help of a full theoretical investigation carried out in parallel with this work~\cite{Coquinot2022}, we determine that the current generation is mainly due to droplet displacement exciting a phonon wind within the graphene sample, its magnitude being tuned by a subtle interplay \MLcorr{between electron-phonon interaction and liquid solid friction.} 

The paper is organized as follows. In Section II we describe our experimental setup and the procedure for a typical experiment. Section III summarizes our quantitative results for the flow-induced electronic current across a wide range of samples and conditions. In Section IV, we outline our theoretical model, and Section V establishes our conclusions.}

\section{Experimental procedure}

\begin{figure*}
\centering
\includegraphics[width=15cm]{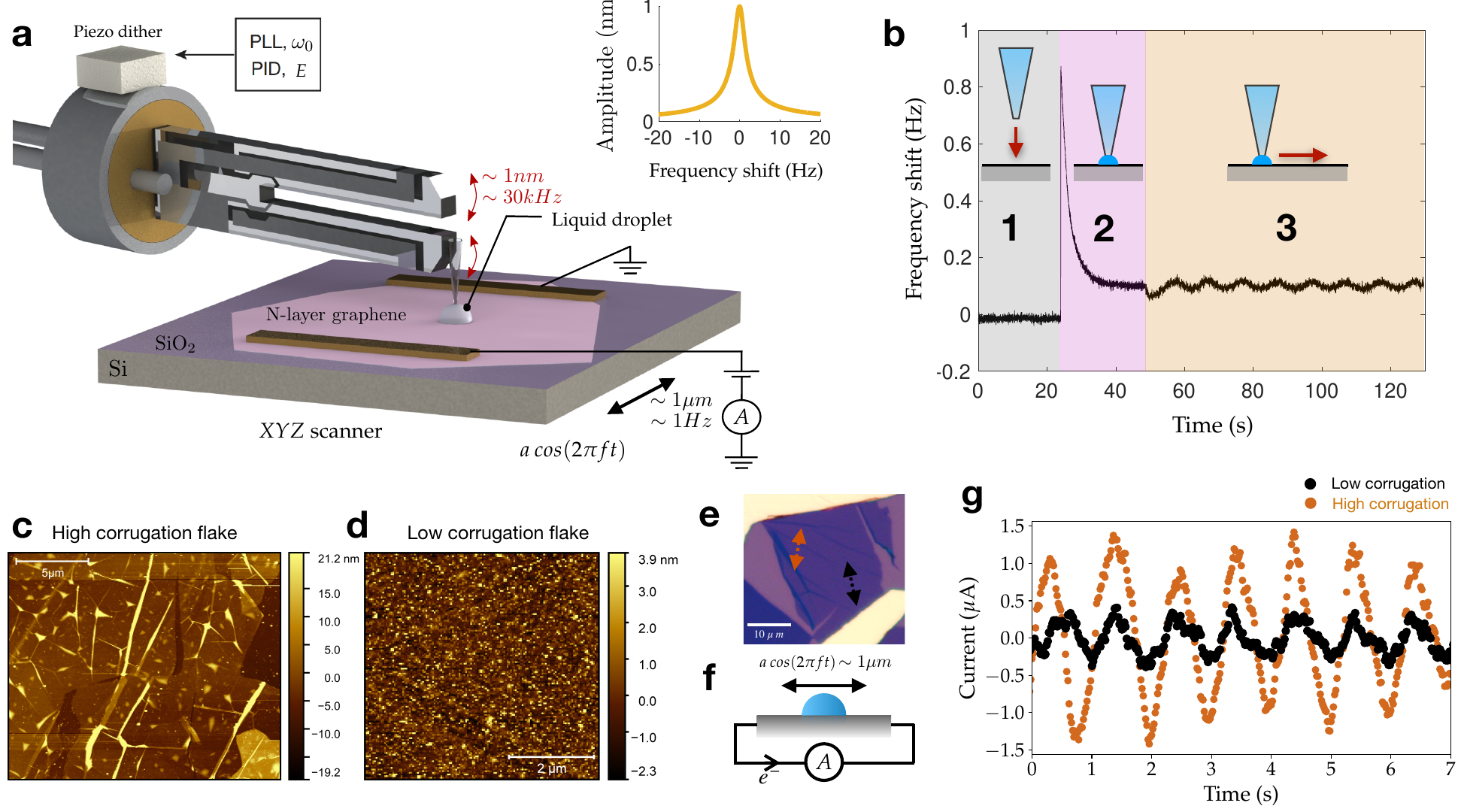} 
\caption{\label{fig1} \textbf{Experimental set-up.} \REF{\textbf{a,} Schematic of the experiment. A quartz capillary of micrometric tip diameter is filled with liquid and glued on the lower prong of a millimetric quartz tuning fork. Once the tip is in contact with the graphite substrate, a liquid droplet is formed at its extremity. The scanner allows for droplet relative motion on the graphite flake in the XY-plane, while current is measured between both gold electrodes. An additional electric potential drop $\Delta V$ can be imposed between the electrodes. \textit{Inset}: The resonance curve of the tuning fork whose frequency shift is used to deposit the droplet and maintain a gentle mechanical contact. }\textbf{b,} Frequency shift response of the AFM during a typical experiment, indicating the stiffness of the pipette-sample contact. First, the pipette is brought into contact with the sample (1). After the contact, a slow relaxation (2) indicates the formation of the droplet. Finally, the sample is put into horizontal motion with respect to the droplet while a PID controller keeps a constant frequency shift (3). \textbf{c,} AFM image of a transferred multilayer graphene sample (3.6 nm thick) exhibiting a high density of wrinkles. \REF{\textbf{d,} AFM image of a flat (un-transferred) multilayer graphene sample (1.5 nm-thick)}. \textbf{e} Optical microscope image of a multilayer graphene sample : in the orange region, some wrinkles increase the roughness compared to the black region. \textbf{f,} Schematic of droplet motion. \MLcorr{ \textbf{g,} Current generated by the droplet motion on a few layer graphene sample under a 1.2 V voltage drop (DC component is substracted) in the orange and black regions shown in panel \textbf{e}. A current enhancement by wrinkles is observed.}}
\end{figure*} 

We deposited a thin graphite flake (1 \MLcorr{monolayer} to 70 nm in thickness) onto a Si/SiO$_2$ substrate. Two gold electrodes, separated by a distance of typically $\sim 10\mu$m and connected to a low noise current amplifier, allowed us to measure the electric current through the flake. The sample was placed in a tuning fork atomic force microscope (AFM), that was specifically designed for controlled liquid deposition, see Figure 1\textbf{a}. The liquid was introduced into a quartz capillary, which was then glued to the quartz tuning fork (see Figure S1 \textbf{a,b}). A piezo-dither induced a nanometric oscillation of the tuning fork at its resonant frequency ($f_0$ $\approx$ 32 kHz), a typical resonance curve is shown in the inset of Figure 1\textbf{a}. By monitoring precisely the tuning fork oscillation's phase and amplitude as the pipette tip approaches the sample, we achieved a fully controlled contact between the carbon surface and the pipette's aperture, leading to the formation of a liquid capillary bridge. The capillary bridge then evolved into a droplet, of radius fixed by the outer diameter of the tip (of micrometric size) \cite{Fang2006, Fabie2012}. This novel experimental set-up allows to deposit very viscous liquid at nano and micro scale over a broad range of substrates \cite{hummink}. We used a room temperature ionic liquid (BMIM-PF6), \MLcorr{glycerol}, and a neutral and apolar silicone oil (Polyphenyl-methylsiloxane), \MLcorr{with respective viscosities of 0.3 Pa$\cdot$s, 1 Pa$\cdot$s and 0.1 Pa$\cdot$s, respectively}, and with very small vapor pressures, allowing us to safely neglect evaporation. The continuous monitoring of the contact with both the in-situ optical microscope and the AFM frequency shift signal allows us to ensure that the droplet remained hooked to the tip, sliding over the carbon surface. 

\REF{In a typical experiment, the droplet is put into oscillatory motion, horizontally on the carbon surface by a piezo-scanner with a \MLcorr{micrometric amplitude} at a frequency of the order of 1 Hz (see Figure 1\textbf{a,f}).} An electric current through the carbon surface is then measured at the droplet oscillation frequency (Figure 1\textbf{e-g}) \MLcorr{for all three liquids}. As shown on Figure 1\textbf{a}, our set-up enables to further impose \REF{a constant potential drop $\Delta V$ between the drain and the grounded source}.  We report electric currents ranging from nano to micro ampere induced by a droplet moving at a few micrometers per second \REF{in samples whose resistance ranges from 300\textrm{$\Omega$} to 2\textrm{$k \Omega$}}. To allow for a comparison with other current generation experiments, we define the current density as the AC-current amplitude divided by the diameter of the droplet (of order $1 \rm \mu m$) : $j=I/(2R_{\rm drop})$ typically in the tens \MLcorr{to hundreds} of $\rm{nA/\mu m}$ range.

Several benchmark measurements allowed us to  ensure that the current generation is due to the specific interaction of the liquid droplet with the few-layer graphene (see Appendix C). We checked that no alternating current is recorded when the deposited droplet oscillates over the SiO$_2$ substrate between the electrodes, \ML{outside the multilayer} graphene substrate. We further controlled that no current is generated when the pipette is moving above the droplet or even when a tungsten tip is used instead of the liquid filled pipette. \MLcorr{We also checked that the tuning fork's oscillation amplitude and the solid contact between the pipette and the graphene have no influence on the results.} Altogether, we can safely conclude that the electro-fluidic current is due to the displacement of the liquid droplet on the carbon surface.

\section{Electro-fluidic conductivity}
\subsection{Scaling behavior}

On Figure 2\textbf{a}, we report results of several samples (with thickness varying between 1.7 and 8.2 nm), where an electro-fluidic current is measured \MLcorr{under an ionic-liquid droplet}. We observe that for a fixed oscillation frequency $f$ and bias voltage $\Delta V$, the amplitude of the AC generated current increases linearly with the droplet oscillation amplitude $a$, or equivalently, with the droplet peak velocity ($v=2 \pi f a$). Taking advantage of the demonstrated linear dependency of the current density with respect to the droplet peak velocity, we define the electro-fluidic conductivity of our samples as 
$$\sigma_{\rm ef} = {j\over v},$$ 
quantifying the cross-coupling between electronic current $j$ and liquid flow $v$. Among the five samples presented here, we show an electro-fluidic conductivity ranging from 2.8 to 21.6 $\rm{nA.s/\mu m^2}$ for \REF{a micrometric ionic liquid droplet.  We stress that non-ionic liquids like silicone oil and glycerol yield an electro-fluidic conductivity of the same order of magnitude as ionic liquids.}\ML{ For example, see Figure S1\textbf{g} where we obtain $\sigma_{\rm ef} \sim 2 \rm{nA.s/\mu m^2}$ with silicone oil.}

\begin{figure*}
\centering
\includegraphics[width=15cm]{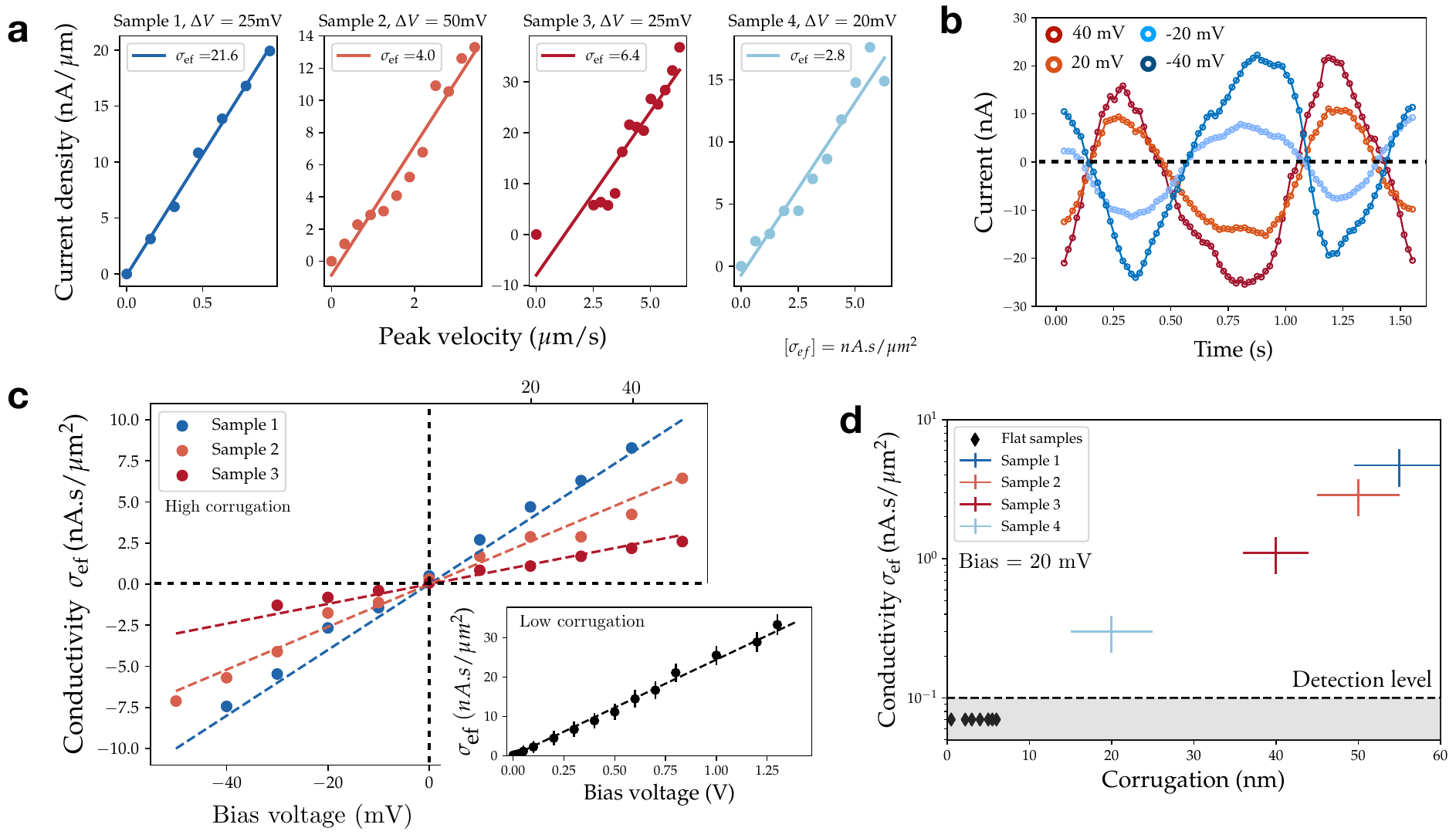} 
\caption{\label{fig3} \textbf{Generation of a droplet motion-induced electronic current.} \textbf{a,} Current density (nA/$\mu m$) as a function of fluid velocity $v=2 \pi f a$ for 4 different samples. Each curve is measured at a bias $\Delta V$ indicated in the corresponding plot. 
The multi-layer graphene subtrates have thickness 3.3, 1.7, 3.6 and 8.2 nm, respectively and their corresponding surface roughness measured using AFM are $\{55,50,40,20\}$ nm respectively. 
\textbf{b,} Time-dependent current under droplet oscillation for various voltage bias $\Delta V$. 
\textbf{c,} Effect of the voltage bias $\Delta V$ on the electro-fluidic conductivity $\sigma_{\rm ef}=j/v$, showing a linear dependency.  The dotted lines are guides to the eye. \MLcorr{\textit{Inset :}Electro-fluidic conductivity on low corrugation few-layer graphene under a bias voltage up to 1.3 V.}
\textbf{d}, Electro-fluidic conductivity $\sigma_{\rm ef}$ (at $\Delta V = 20$ mV)  versus the substrate corrugation, measured using AFM images. The black diamonds correspond to current measurements below the detection level.
}\end{figure*}

To gain insight into the mechanism driving this electro-fluidic current and having in mind the strong \NK{dependence of the graphene electronic density on the local potential} \cite{Neto2009}, we tune the bias voltage $\Delta V$ between the electrodes and measure its influence on the electro-fluidic current. We observe \NK{that the amplitude of the alternating current scales linearly with $\Delta V$, as} highlighted in Figure 2 \textbf{b, c}. \NK{We observe a small or negligible current at zero potential drop and we note that the current is phase-shifted by $\pi$ between positive and negative values of the bias voltage (this is represented by the sign change in the amplitude).} Finally, the slope  $\sigma_{\rm ef}/\Delta V$ is in the range of  0.05 to 0.2 $\rm{nA.s/\mu m^2/mV}$, see Figure 2\textbf{c}, with a slight variation across samples.  

\subsection{Wrinkles as current amplifiers}
An unexpected feature of the experimental results is the sharp distinction among carbon samples regarding the amplitude of the the electro-fluidic conductivity. \NK{Across all samples}, 
we reveal a correlation between the electro-fluidic conductivity and the existence of surface corrugation. Indeed, the transfer of graphene can induce folds and wrinkles on thinnest samples whereas the thicker ones present a flat surface (see Appendix A, Figure 1\textbf{c} and S2\textbf{b}). The thinner flakes, being softer and more flexible than the thicker ones, are more likely to exhibit wrinkles \cite{Liu2011, Deng2016}. In order to disentangle the role of wrinkles from that of sample thickness, we implemented an alternative fabrication method in which all transfer steps are avoided and few-layer graphene flakes were directly exfoliated on the Si/SiO$_2$ wafer.
This allowed us to obtain sub-10 nm, {un-transferred} and 
\textit{smooth} graphene samples with a uniform wrinkle-free surface to compare with the \textit{transferred} and \textit{crumpled} flakes of similar thickness (see Figure 1\textbf{c,d}) showing a much larger surface corrugation (see Appendix D for a Raman analysis of wrinkles). As shown on Figure 2\textbf{d}, droplet motion on smooth samples results in no detectable current \MLcorr{under a small 20 mV bias voltage}. Conversely, crumpled flakes of the same thickness have a high wrinkle density and show a strong electro-fluidic conductivity, in the $\rm{nA.s/\mu m^2}$ range. \MLcorr{The electro-fluidic current in flat un-transferred samples can nonetheless be measured under strong bias voltage (see inset of Figure 2\textbf{c}). To further establish the role of wrinkles, we compare the electro-fluidic current in two regions of the same sample showing respectively a high and low corrugation. The results shown in Figure 1\textbf{e,g} demonstrates clearly the current-enhancing role of the wrinkles.} As summarized in Figure 2\textbf{d}, our results on many different samples reveal that \MLcorr{a strong corrugation in the form of wrinkles strongly enhances the electro-fluidic current.} 

Further evidence of the crucial role of wrinkles is provided by the dependence of the generated current on the droplet oscillation direction on corrugated samples, see Figure S2 \textbf{c,d}. As expected, the current amplitude is 180$\degree$ periodic with respect to the angle between the droplet trajectory and the electrodes. However, \MLcorr{on corrugated samples,} the maximal amplitude is not necessarily reached when the droplet oscillation is perpendicular to the electrodes, and the direction yielding the maximum current strongly depends on the position on the sample's surface. We argue that this spatial and directional variability is related to the random distribution in height, position and direction of the wrinkles on the carbon surface. This further highlights the role of wrinkles as current amplifiers.  Let us note for completeness that the data displayed on Figure 2 are obtained along the direction with maximal amplitude.

\section{Phonon drag mechanism}
Summarizing our experimental results, we have shown that the motion of a micrometric droplet on a few-layer graphene surface generates a strong electro-fluidic current in the flake, which is proportional to the velocity of the droplet, as well as to the bias voltage applied between the electrodes. A remarkable feature is the strong dependence of the magnitude of the generated current on the surface corrugation of the flake. 

\begin{figure*}
\centering
\includegraphics[width=15cm]{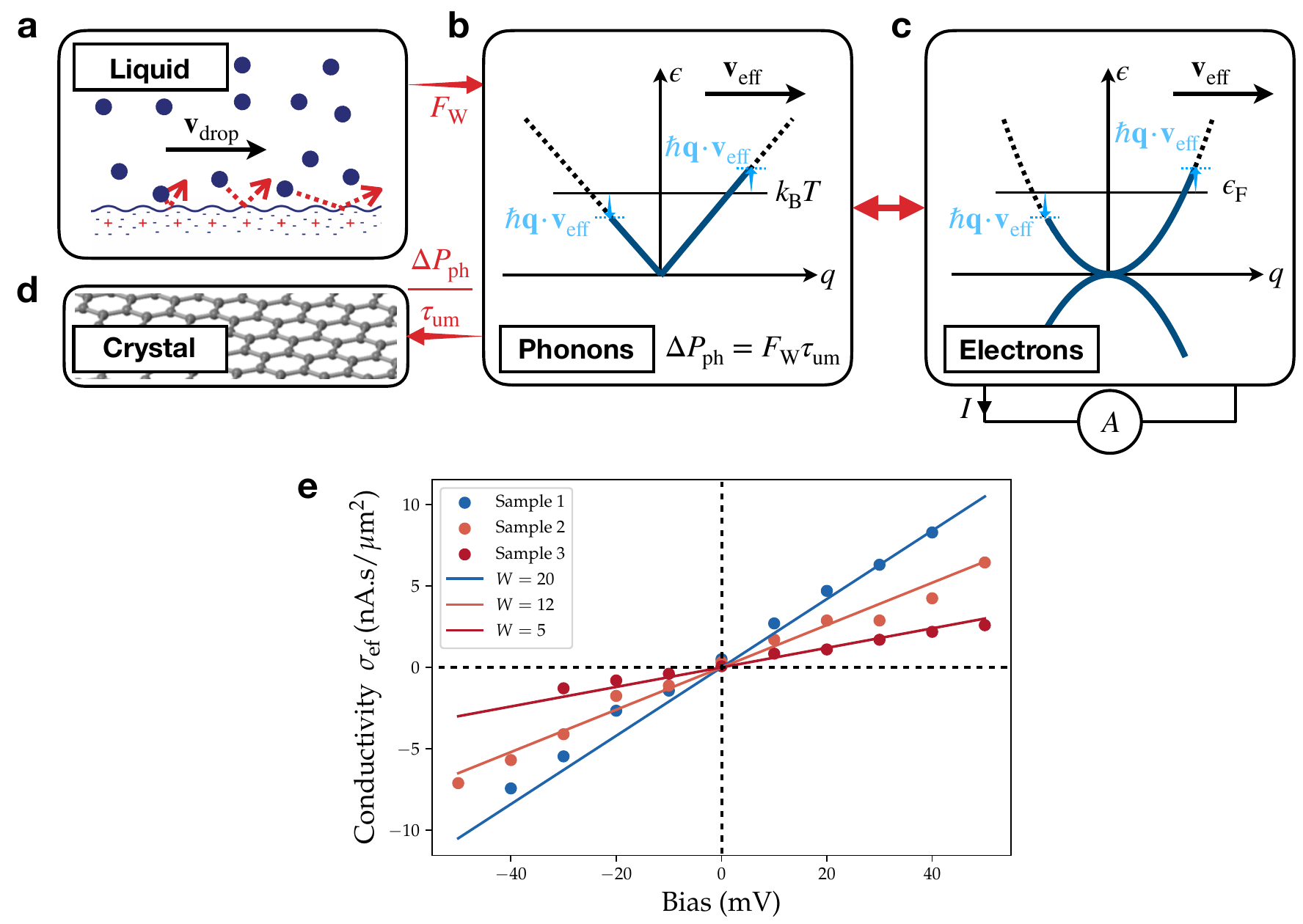} 
\caption{\label{theory} \textbf{Phonon drag mechanism.} The flowing liquid \textbf{a} exerts a friction force $F_{\rm W}$ on the graphite sample. This force is related to a momentum transfer to the phonons, that generates an asymmetry in the phonon distribution \textbf{b} which is biased in the direction of the liquid flow.  Finally this asymmetry propagates to the electronic distribution \textbf{c} by electron-phonon interaction. This asymmetry in the electronic distribution translates into the generation of a net electric current, whose sign depends on the nature of charge carriers at the Fermi level (electrons or holes).  We stress that this is an out of equilibrium phenomenon in which most of the momentum is relaxed to the crystal \textbf{d} by the Umklapp process with a rate $1/\tau_{\rm um}$.  Finally, on \textbf{e}, we compare the theoretical derivation of the electro-fluidic conductivity $\sigma_{\rm ef}$ versus the bias voltage $\Delta V$.  We stress that the experimental data is the same as in Figure 2\textbf{c} and that the solid lines are plotted using  Eq.~\eqref{quantitative} assuming the bulk graphite effective mass $m=0.1m_{\rm e}$ and $W$ the wrinkling number as the only fitting parameter chosen between 5 and 20. This wrinkling number characterizes the enhancement of Stokes friction due to wrinkles on the graphene surface.}
\end{figure*}

While our results echo previous reports of electric potential differences induced by water flow in carbon-based nanomaterials \cite{Kral2001, Gosh2003, Cohen2003, Persson2004, Zhao2008, Rabinowitz2020,Dhiman2011, Yin2012, Lee2013,Kuriya2020, Yin2014, Park2017}, the nanoscale dimensions of our experiment put us in position to disentangle various effects at the root of electronic current generation, and obtain insights into the molecular mechanism underlying the phenomenon. Previously, several mechanisms have been invoked to rationalize electro-fluidic currents, including streaming potential effects \cite{Cohen2003}, adsorbed/desorbed ion hopping \cite{Persson2004, Dhiman2011}, Coulomb drag \cite{Gosh2003, Rabinowitz2020} or charging/discharging of a pseudo-capacitance \cite{Yin2014, Park2017}. These mechanisms \NK{rely on} Coulomb interactions between the polar liquid -- that carries molecular partial charges -- and the carbon surface as the main pathway for momentum transfer from the liquid to the electrons. \NK{However, the strong impact of surface corrugation suggests that, in the present case, current generation requires 'mechanical' transfer of momentum from the fluid to the charge carriers. } \MLcorr{This hypothesis is reinforced by the relative independence on the liquid nature as ionic liquids, silicone oil and glycerol all yield comparable currents. Liquid-solid Coulomb interactions are therefore not instrumental to the underlying mechanism.}

This conclusion led us to consider a phonon-drag-based mechanism, as first proposed theoretically in ref.~\cite{Kral2001}. Phonon drag relies on hydrodynamic (viscous) friction at the solid-liquid interface, which transfers momentum to the solid's phonons, thereby exciting a 'phonon wind'. Through the electron-phonon interaction, the phonon wind 'blows' on the Fermi sea, so that electrons near the Fermi surface are dragged along the direction of the liquid flow, hence producing an electric current (see Figure 3\textbf{a}). A complete quantum description of phonon drag, within the non-equilibrium Keldysh framework, \NK{is developed in the accompanying theoretical paper}~\cite{Coquinot2022}. Here, we give only a semi quantitative description that allows us to capture the main physical ingredients. 

First, assuming a Couette flow within the liquid droplet, the hydrodynamic friction force is $\textbf{F}_0 \sim \eta \mathcal{A}_{\rm drop}  \textbf{v}_{\rm drop}/h_{\rm drop}$, $v_{\rm drop}$ is the drop velocity, $\eta$ its viscosity,  $\mathcal{A}_{\rm drop}$ its surface and $h_{\rm drop}$ its height
(here typically, $\eta=0.3\, \textnormal{Pa}\cdot\textnormal{s}$ and $h_{\rm drop}\approx 1\, \mu\textnormal{m}$). 
In line with observations, wrinkles are expected to enhance this hydrodynamic friction force. In a simple picture, wrinkles are modelled as protrusions 
on which the flowing liquid exerts a Stokes force that scales (per wrinkle) as $\textbf{F}_{\rm W} \propto \eta h_{\rm W} \textbf{v}_{\rm drop}$, where $h_{\rm W}$ is the typical height of the wrinkles ($h_{\rm W} \sim 10-30 ~\rm nm$). One then obtains a total friction force $\textbf{F}_{\rm W} \sim \textbf{F}_0 (1+W)$, where we introduced the dimensionless wrinkling number, $W= 3\pi n_{\rm W}h_{\rm W}h_{\rm drop}$ ($n_{\rm W}$ is the wrinkle density) to account for the increase in viscous friction due to surface corrugation.  \REF{We emphasize that our approach relies on a momentum balance between the fluid and the solid. Hence the analysis does not depend crucially on the specific nature of the liquid flow. The Couette flow is merely a first approximation to the flow which is sufficient to obtain an estimate of the friction force.}

\NK{A simplified sketch of our theoretical model} is shown in Figure 3 \textbf{a-d}. The hydrodynamic friction force transfers momentum from the liquid (Figure 3\textbf{a}) to the solid's acoustic phonons, which have a dispersion relation $\epsilon_{\mathbf q}=\hbar q c$, $c$ being the phonon (sound) velocity ($c \approx 2\times 10^4 ~\rm m\cdot s^{-1}$ \cite{Nika2009,Ochoa2011,Cong2019}). The phonons relax mainly by transferring their momentum to the \NK{underlying substrate} (Figure 3\textbf{d}) through Umklapp processes \cite{Nika2009},
on a typical timescale $\tau_{\rm um}\sim$ 10 ps \cite{Klemens1994}.  As sketched on Figure 3\textbf{b}, the constant influx of momentum \NK{can be modeled by a 'Doppler shift' in the phonon distribution}, according to  $\epsilon_{\mathbf q} \mapsto \epsilon_{\mathbf q} - \hbar \mathbf{q} \cdot \mathbf{v}_{\rm ph}$. Here $\mathbf{v}_{\rm ph}$ is the velocity of the phonon wind: it is determined by balancing the momentum fluxes in and out of the phonon system, according to $\tau_{\rm um} \textbf{F}_{\rm W}/\mathcal{A}_{\rm drop}  =\Delta\textbf{P}_{\rm ph}/\mathcal{A}_{\rm drop} = \int \frac{\mathrm{d} \mathbf{q}}{2\pi} \, \hbar\mathbf{q} \,n_{\rm B} (\hbar q c-\hbar\mathbf{q}\cdot \textbf{v}_{\textnormal{ph}})$, with $n_{\rm B}$ the Bose-Einstein distribution and $\mathbf{q}$ is the phonon wavevector, that we assume two-dimensional. This yields
\begin{equation}
 \textbf{v}_{\textnormal{ph}} \simeq \frac{2\pi}{3 \zeta(3)} \frac{\hbar^2 c^4}{(k_BT)^3}\times (1+ W)\eta\frac{\textbf{v}_{\rm drop}}{h_{\rm drop}} \times\tau_{\rm um}  	
 \label{vph}
 \end{equation}
 Then,  $v_{\textnormal{ph}} \approx 9\cdot 10^4 \times (1+ W) \times v_{\rm drop}$.  
 
As the electrons scatter on the phonons, \NK{they acquire an average 'wind' velocity $\mathbf{v}_{\rm e}$. If the direct liquid-electron Coulomb interaction is negligible, as considered in the simplified model of Figure 3\textbf{a-d}, the electronic wind velocity aligns to the phonon wind velocity ($\mathbf{v}_{\rm e} \approx \mathbf{v}_{\rm ph}$), regardless of the form of the electron-phonon interaction~\cite{Coquinot2022}: this is a good approximation in the case of an apolar liquid. If the liquid is polar, our full theoretical analysis~\cite{Coquinot2022} shows that the electronic wind velocity is in fact given by a linear combination of the phonon wind and interfacial flow velocities:  
\begin{equation}\label{ve}
\textbf{v}_{\rm e}=\frac{\tau}{\tau^{\rm e/ph}}\textbf{v}_{\rm ph}+\frac{\tau}{\tau^{\rm e/h}}\textbf{v}_{\rm \ell}.
\end{equation}
 Here $\tau^{\rm e/ph}$ and $\tau^{\rm e/h}$ are the typical electron-phonon and electron-liquid scattering times, respectively, and $\tau^{-1} = (\tau^{\rm e/ph} )^{-1} + (\tau^{\rm e/h})^{-1}$ is the total electron scattering rate near the Fermi surface. According to eq.~\eqref{vph}, $v_{\ell} < v_{\rm drop} \ll v_{\rm ph}$, but, at the same time, the electron-phonon and electron-liquid scattering times have a similar order of magnitude~\cite{Coquinot2022}. Therefore, the electronic wind induced by a polar liquid is given by $\textbf{v}_e \approx \textbf{v}_{\rm ph}/2$: the liquid-electron Coulomb interactions actually contribute to reducing the flow-induced current. Accounting for the electronic wind amounts, again, to a Doppler shift of the electron dispersion in momentum space: $\epsilon_{\mathbf q} \mapsto \epsilon_{\mathbf q} - \hbar \mathbf{q} \cdot \mathbf{v}_{\rm e}$.}

Assuming for concreteness that the liquid flows to the right, this roughly means that in an energy window of width $\hbar k_{\rm F} v_{\rm eff}$ around the Fermi level ($k_{\rm F}$ is the Fermi wavevector), the electrons move to the right at the Fermi velocity $v_{\rm F}$. The corresponding current density reads 
\begin{equation}
\textbf{j} \approx e v_{\rm F}  \times N(\epsilon_{\rm F}) \times [\hbar k_{\rm F} \textbf{v}_{\rm e}], 
\label{jth}
\end{equation}
where $N(\epsilon_{\rm F})$ is the density of states at the Fermi level $\epsilon_{\rm F}$.  \NK{This simple expression is in fact consistent with the low-temperature limit of our full derivation~\cite{Coquinot2022}, and provides enough quantitative accuracy for a comparison with experimental results. }
\NK{Equation ~\eqref{jth} predicts} a current density proportional to the droplet velocity, to the liquid viscosity and to the wrinkling number W, in excellent agreement with the experiment. Moreover, the current is independent (\NK{up to a factor of at most 2 that depends on the liquid polarity}) of the chemical nature of the liquid, \NK{consistently with a similar amount of current being generated by ionic liquid, glycerol and silicone oil droplets. }

Although the graphene electronic bands have a linear dispersion, the electronic structure our multilayer samples of few-nanometer thickness is expected to be better described by two touching parabolic band with effective mass $m \sim \pm 0.1 m_e$ ($m_e$ being the free electron mass) \cite{Gruneis2008, Guinea2008, Charlier1992}: $\epsilon_{\mathbf{q}} = \pm \hbar^2 q^2 / 2 m$.  The density of states is then independent of energy: $N(\epsilon) = N(\epsilon_{\rm F}) = m/\hbar^2$.  Altogether, we obtain the electro-fluidic conductivity as
\begin{equation}\label{prediction}
\sigma_{\rm ef} = e \frac{m}{\hbar^2}\epsilon_{\rm F} \times \frac{v_{\textnormal{e}}}{v_{\rm drop}}.
\end{equation}

The prediction in Eq.~\eqref{prediction} can now be compared quantitatively with experimental results. \NK{ Experimentally, the bias voltage modifies the local potential at the droplet position, setting the Fermi level $\epsilon_{\rm F} \approx e \Delta V /2$}, so that 
\begin{equation}\label{quantitative}
\sigma_{\rm ef}  \approx  0.1\times m/m_{\rm e}\times (1+W)\times \Delta V  \,[\textnormal{nA.s/}\mu\textnormal{m}^2]
\end{equation} 
with $\Delta V$ in mV. This estimate quantitatively reproduces the experimentally measured conductivity of a few nA.s/$\mu$m$^2$ as well as its linear scaling with the bias voltage. 
\NK{We note as well that the current is expected to change sign if the charge carriers are holes instead of electrons.} This explains why the generated current flips sign along with $\Delta V$: for positive bias, the charge carriers are electrons in the conduction band, whereas for negative bias, the charge carriers are holes in the valence band. \NK{In Figure 3\textbf{e}, we compare our theoretical prediction with the experimental results for $\sigma_{\rm ef}$ and observe indeed quantitative agreement, the only fitting parameter being the wrinkling number W, chosen between 5 and 20.} \REF{Having developed the qualitative picture of phonon drag induced current and provided a quantitative comparison to experiments, we now discuss briefly the role of wrinkles as current amplifiers.  In our drag-increasing-protrusion picture, wrinkles enhance the momentum transfer between the fluid and solid by actively increasing the interfacial friction force.  In addition to this effect, it is known that wrinkles induce large strain gradients  \cite{Neumann2015} (see Appendix D and figure S2g-j for Raman maps), locally enhancing the electron-phonon scattering and thus the momentum transfer from phononic to electronic distribution.  While our interpretation provides a quantitativaley consistent explanation, one cannot rule out such enhanced scattering effects and they may indeed contribute to current enhancement pointing out the important role of local structure to phonon drag phenomenon.}

\NK{Altogether, our theoretical description strongly supports the hypothesis that the experimentally observed current is due to the graphene charge carriers being dragged by a phonon wind induced by the liquid flow. In the accompanying theoretical paper~\cite{Coquinot2022}, we further predict that such a mechanism leads to a \emph{quantum feedback} phenomenon, where the electrons return part of the momentum acquired from the phonon wind back to the liquid: this is the reason why the electron wind velocity cannot align to phonon wind if the flowing liquid is polar (see eq.~\eqref{ve}). The quantum feedback leads to an apparent reduction of the hydrodynamic friction; measuring this 'negative friction' is beyond the scope the present experiments. At the relatively low electronic densities achieved in our samples, we would expect a friction reduction of at most a few percent~\cite{Coquinot2022}. Specific friction measurements in samples with higher electronic density of states, where the effect is expected to be macroscopic, will be the subject of future work. }

\section{Conclusions}

We report the generation of electronic current in multilayer graphene induced by the controlled motion of a liquid droplet. Our results suggest that the current generation is mediated by hydrodynamic momentum transfer from the flowing liquid to the phonons in the solid. Our specifically tailored experimental system is particularly well-suited to evidence this phonon drag mechanism. Indeed, the combination of a local micrometric flow on graphene samples of tunable corrugation with the use of highly viscous neutral liquids allows the phonon drag to dominate current generation over other possible mechanisms. The versatility of our setup enables us to tune the fluid velocity as well as the solid's electronic properties using the bias between the electrodes to fully characterize the phenomenon. 

The experimental results were rationalized with a new dedicated theoretical model. \NK{The experimental evidence for the basic phonon drag mechanism is promising for the observations of further predictions of our theory, including the quantum feedback mechanism and 'negative friction'}.   Our findings pave the way for the active control of fluid transport at the nanoscale, by harnessing the interplay between molecular liquid flows and collective excitation within the solid.  \REF{They further stimulate the development of small footprint flow sensing devices which could allow simultaneous measurement of interfacial velocity and viscosity.} Finally, the coupling of electronic and micro-fluidic currents bears the promise of ubiquitous and environment-friendly energy production, as demonstrated recently in 
transpiration-driven electrokinetic power generators \citep{Bae2022}.


\section*{Appendix A : Graphene-based device’s micro-fabrication}
\subsection{Transferred samples}

The HOPG crystals are purchased from HQ Graphene. Each bulk crystal is mechanically exfoliated and flakes are first randomly deposited on separate 280 nm-thick Si/SiO2 substrates.  Graphene flakes are then carefully selected for their thickness (ranging from 1 nm to 70 nm), homogeneity and size (> 15 $\mu$m).  The graphene flakes are transferred thanks to the “hot pick-up” technique \REF{known to produce clean and contamination free surfaces}  \citep{Pizzocchero2016, Rastikian2019} in a 100 x 100 $\mu$m$^2$ area in the middle of a 2x2 mm$^2$ pre-patterned substrate with Au electrodes. 
The devices are then fabricated by defining two parallel electrical contacts over the graphene flake by e-beam lithography using a double layered PMMA process: a 50 K lower layer of PMMA and a 950 K upper layer of PMMA both spin-coated at 4000 RPM. The spacing between the two contacts is set to at least 10 $\mu$m for the AFM experiment. The metallic contacts are subsequently deposited under high vacuum (a few 10-7 mBar) by e-beam evaporation.  5 nm of Ti covered by 45 nm of Au are deposited at very low evaporation rates (0.1 nm/s). The last fabrication step is a lift-off process in which the remaining resist is dissolved in an aceton bath for 20 min. The samples are finally rinsed into isopropanol for 2 min. This fabrication method involves a transfer of the graphite flake and thus yields a high density of wrinkles on thinner flakes as measured by AFM (\textit{cf:} Figure 1\textbf{c}). Thicker flakes on the other hand do not show any tranfer-induced wrinkles (\textit{cf:} Figure S2\textbf{b}).

\setcounter{figure}{0}
\captionsetup{labelformat = suppl}
\begin{figure*}
\centering
\includegraphics[width=15cm]{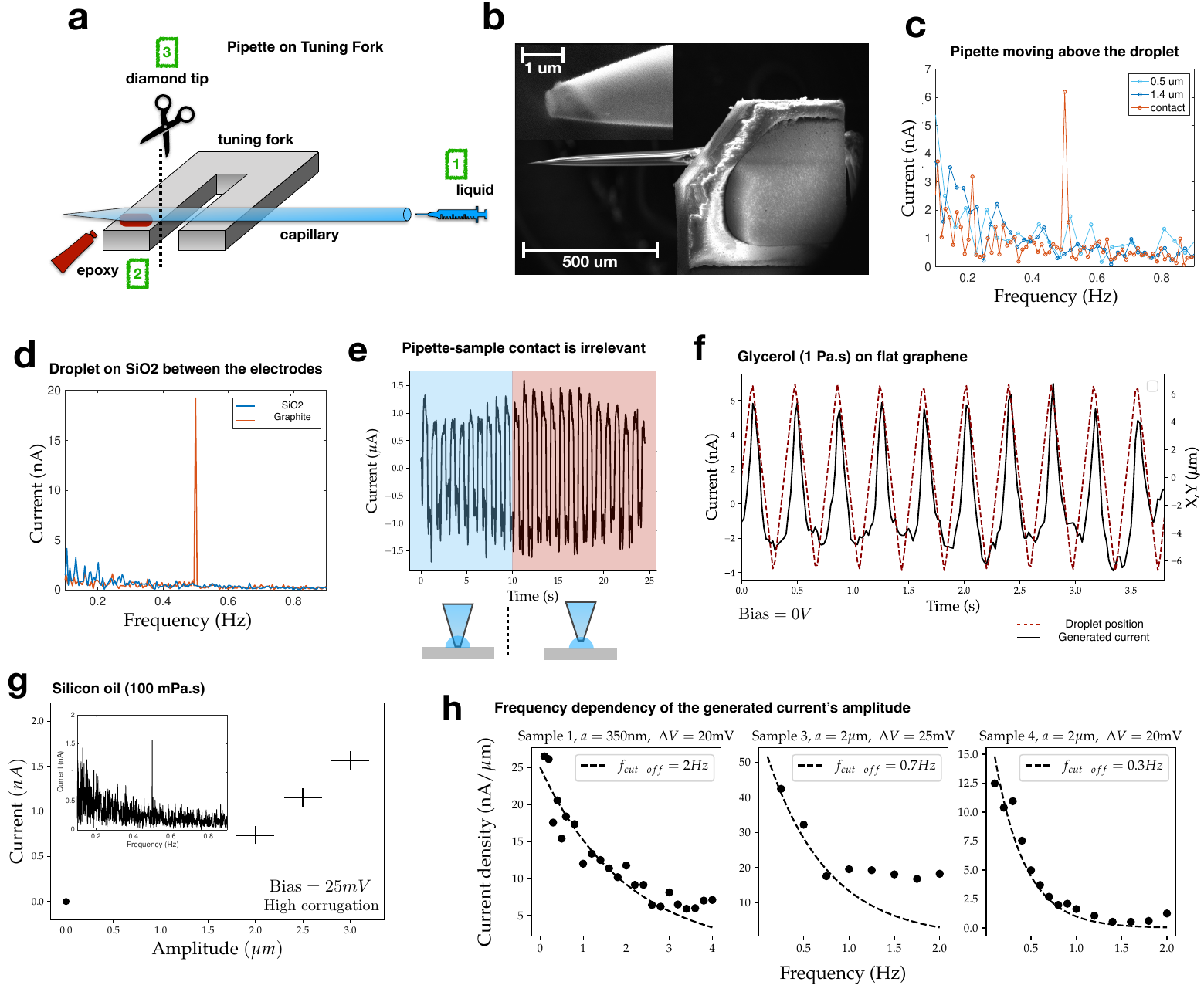} 
\caption{\label{figSI1}  \textbf{Experimental techniques and controls} \textbf{a} Steps for the preparation of the tuning fork and its probe: Firstly, the pipette is filled with liquid. Then, it is glued with epoxy on the side of one tuning fork prong.  Finally, the cylindrical part of the capillary is cut free from its extremity and from the tuning fork with a diamond tip. \REF{\textbf{b} Scanning Electron Microscopy images of the lower prong of a millimetric tuning-fork, and of the extremity of the pipette glued on it (\textit{inset}).} \textbf{c} Pipette is set to move above the droplet which remains at rest on the sample. We see that droplet motion is necessary. \textbf{d} The droplet generates current on graphene but not on SiO2. \MLcorr{\textbf{e} Generated current under droplet motion. At $t=10s$, the pipette is retracted by 300nm far from the solid contact but remains in the droplet and maintains its motion. We show here that the pipette-graphene interaction is of no importance in the current-generation mechanism.} \REF{\textbf{f} \textit{With glycerol} : electro-fluidic current \textit{versus} time with a 10$\mu$m droplet on a flat graphene sample with zero bias under a 10 V back-gate voltage. \textbf{g} \textit{With silicone oil} : generated current amplitude as a function of the silicone oil droplet oscillation amplitude, at 0.5 Hz and \CORR{a bias voltage $\Delta V = 25 mV$.} We extract $\sigma_\textrm{ef}\sim2nA.s/\mu m^2$ (\textit{inset :} 3 $\mu$m amplitude). } \textbf{h} Frequency dependency of the electro-fluidic current at fixed oscillation amplitude. \CORR{An exponential decay of decay frequency $f_{cut-off} \in \left\{ 2,0.7,0.3 \right\} Hz$ is also shown for each sample.}} \end{figure*}

\begin{figure*}
\centering
\includegraphics[width=15cm]{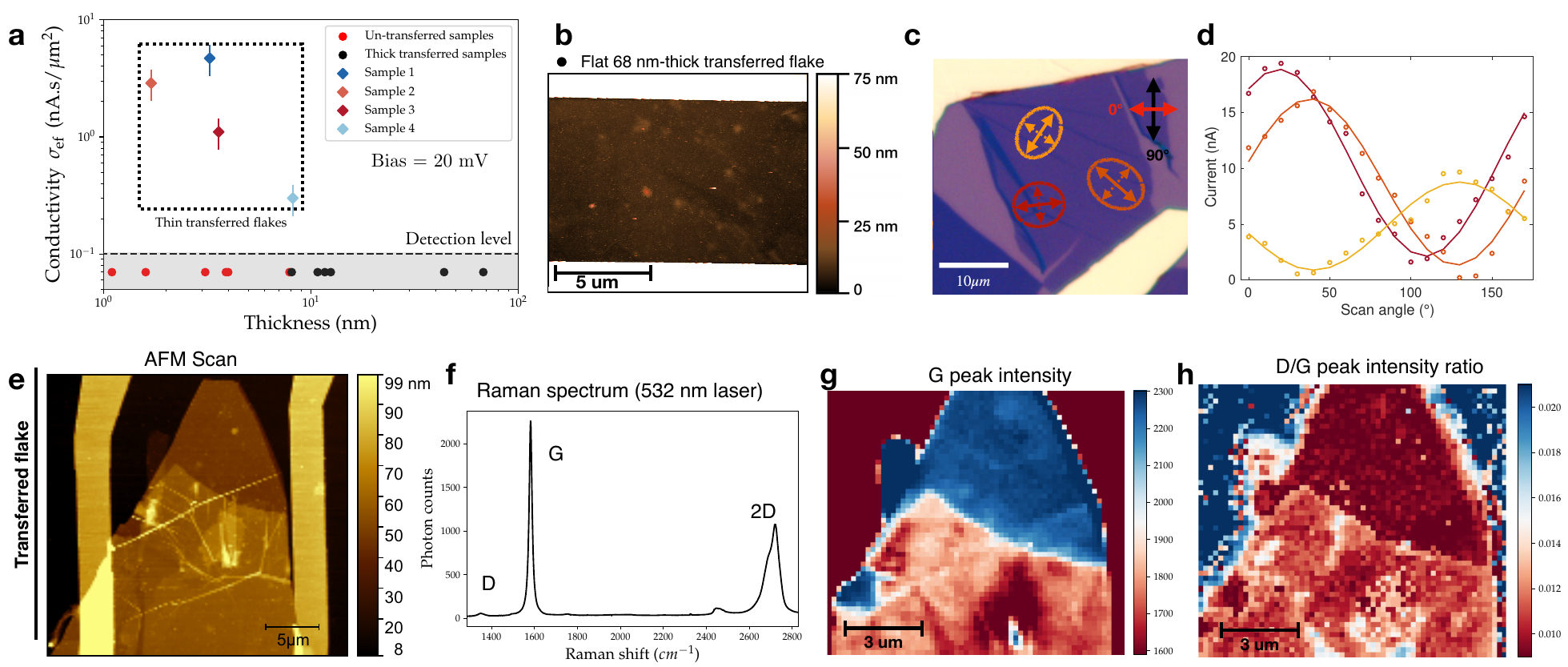} 
\caption[figurename=Figure S]{\label{figSI2}\textbf{Influence of wrinkles on graphene surface} \REF{\textbf{a} Log-log plot of the electro-fluidic conductivity given in $nA.s/\mu$m$^{2}$, as a function of the flake thickness. We clearly show that the fluctuations of electro-fluidic conductivity are dominated by corrugation and not flake thickness (\textit{cf :} Figure 2\textbf{d}). } \textbf{b} AFM image of a flat 68 nm-thick transferred sample. Its large thickness prevents the creation of wrinkles.  \textbf{c} Cartoon of the angular dependency of the generated current. \textbf{d} Generated current (nA) as a function of the scan angle ($\degree$) of droplet motion (at 0.5 Hz frequency and 1 $\mu$m amplitude), around various central positions. Experimental points are pictured by circles, while the solid line is the best sinusoidal fit.  On \textbf{e}, we show an AFM scan of a wrinkled flake with its electrodes on which we performed extensive Raman spectroscopy.  On \textbf{f}, a typical Raman spectrum is shown. The Raman G-peak intensity is displayed on  \textbf{g} and is stronger in the thicker regions. On \textbf{h}, the intensity ratio between the D and G peak is displayed. The low value ($1 \%$) of this ratio indicate a very low level of defects.  Moreover, wrinkles, as opposed to flake edges do not display any local D-peak enhancement. This indicates that wrinkles are indeed folded graphene structures and neither pollution nor defects.  }\end{figure*}

\subsection{Un-transferred samples}
As discussed in the text, wrinkles have a strong influence on viscous flow driven current generation at the carbon surface. To assess the role of wrinkles in current generation, we used a transfer-free fabrication technique to prepare flat multilayer-graphene samples. We evaporated gold electrodes on the SiO2 substrate where the graphene was exfoliated in the first place and thus obtained wrinkle-free samples (Figure 1\textbf{d}) which can be compared to high wrinkle-density (Figure 1\textbf{c}) samples of the same thickness obtained when transferring the flakes before evaporating the contacts. 

\section*{Appendix B: Droplet motion-induced current generation}
\paragraph*{Micro-pipette AFM preparation}
The capillary tip is obtained by locally heating and simultaneously pulling a 10 cm long quartz capillary of inner diameter 0.5 mm and outer diameter 1 mm (World Precision Instruments),  with a Sutter Instruments P-2000 pipette puller.  We obtain tips with an outer diameter of 500 nm \REF{(measured by SEM)}, that can be varied by tuning the pulling parameters.  The pipette is then filled with ionic liquid (Bmim-PF6 of high purity from Merck), with glycerol or with neutral silicone oil (Polyphenyl-methylsiloxane, 100 mPa.s, from Sigma Aldrich).  Once the pipette is filled, its very extremity is glued with epoxy glue on the lateral side of a millimetric quartz tuning fork prong. With a diamond tip, the largest part of the capillary is then cut free from its extremity and from the tuning fork. These steps are summarized on Figure S1\textbf{a}.  Finally, the tuning fork and its probe are examined under Scanning Electron Microscopy, to measure the outer diameter of the pipette extremity and to make sure the previous fabrication steps did not damage the tip, see Figure S1\textbf{b}. 

\paragraph*{Liquid droplet creation and control}
All our experiments are conducted under ambient conditions. We deposit a small amount of liquid and move it over the sample while keeping constant geometrical properties of the droplet as well as mechanical constraints.  As a liquid micro-manipulation tool,  we use the quartz-tuning fork as an autosensitive Atomic Force Microscope (AFM) mechanical oscillator. The tuning fork is excited \textit{via} a piezo-dither at its resonance frequency $f_0$ in the normal mode ($f_0 \approx 32$ kHz), while the amplitude and phase shift of the tuning fork oscillations with regards to the excitation AC voltage are obtained from the piezoelectric current through the tuning fork electrodes. Our home-made AFM is used in frequency-modulation mode : a Phase Lock Loop (PLL) ensures that the tuning fork is systematically excited at its resonant frequency and a PID servo loop maintains a constant oscillation amplitude by tuning the excitation voltage. Interactions with the substrate are detected by the change in the resonance frequency and in the excitation voltage.  As shown on Figure 1 \textbf{a,b},  we monitor the gentle approach \textit{via} the PLL to detect contact with the substrate and the formation of the capillary bridge, and to ensure a constant contact stiffness (frequency-shift) during the droplet motion on the sample. \MLcorr{As shown on Figure S1\textbf{c}, whether the pipette touches the graphene or not has no influence on the induced-current as long as it drives the liquid flow : the liquid is not confined by the pipette. We verified that the tuning fork amplitude and the frequency shift (pipette-sample contact stiffness) have no influence on the generated current.}


\paragraph*{Role of slippage}
\CORR{Considering the existing literature and experiments reported by several group where the slip length on multilayer graphene/SiO2 to be in the range of few nanometers at most \cite{Greenwood2022,Li2022}, we do not expect any slip flow effect to be relevant in our experiment. Indeed the size of the droplet is governed by the shape of the pipette's aperture and always in the micrometer range. Any slippage effect is expected to scale like the ratio between the slip length and the size of the drop: therefore only a correction in the order of 1$\%$ can be induced by slippage.}
\REF{\paragraph*{Definition of $\sigma_\textrm{ef}$}
We normalize the susceptibility of our experiment by the droplet's diameter thus defining the electrofluidic conductivity as a current divided by a velocity times a diameter. Although one could expect a dependency on the droplet's surface, this is not what our theoretical analysis predicts and we opted for this option to ease the comparison with theory. Still, for the data presented on Figure 2, as the droplet is a half sphere 1$\mu m$ diameter, the surface normalized conductivity can be easily recovered as $\sigma_{ef}/0.75 \mu m$.}

\section*{Appendix C : Controls}
To make sure that the measured current is indeed related to the viscous flow of liquid on graphite, we performed extensive controls which we detail here. \REF{Firstly, the cleanliness of the samples is checked with AFM and airborne contaminants are removed by annealing in Ar atmosphere.  We have not seen any impact of this annealing, this suggests that airborne contamination is irrelevant in the current generation mechanism.} 


\paragraph*{Pipette above the droplet}
We checked that no current can be generated if the pipette is not in contact with the liquid droplet but is slightly above instead (see Figure S1\textbf{c}). In this configuration, a liquid droplet is deposited on graphite and the pipette is subsequently retracted and brought a few $\mu m$ above the liquid. Thus, during the oscillation, no motion is induced between the droplet and the sample. This rules out any coupling between high voltage signal driving the scanner and the generated current \REF{as well as a coupling between trapped charges in the quartz and electronic currents. To further rule out the latter effect, we highlight that the screening length in the ionic liquid we used is much smaller (nm, \cite{Lynden2012}) than the droplet's height (\textrm{$\mu m$}).}

\paragraph*{On SiO$_2$}
We also checked that no current can be generated if the droplet moves over the SiO$_2$ surface between the electrodes, see Figure S1\textbf{d}.  These controls show unambiguously that current generation requires the droplet to slide on the graphite surface.

\paragraph*{Pipette graphene contact is irrelevant}
Finally, we checked the irrelevance of solid contact between the pipette and the graphene surface by retracting of 300nm the pipette during oscillation. The mechanical contact is lost but the droplet is still driven by the pipette. We show that the electro-fluidic current remains the same. Overall, our extensive controls show without ambiguity that a liquid droplet motion on graphene generates electronic current in the latter.

\paragraph*{Ionic or neutral nature of the liquid is irrelevant}
We show on Figure S1\textbf{f,g} that using glycerol or silicone oil droplets, we can generate an electro-fluidic current of the same order of magnitude as when using ionic liquids. With silicone oil especially, on panel \textbf{g}, the droplet size, bias voltage and velocity are comparable to results from Figure 2 and the electro-fluidic current is as well. These observations are a very strong argument for a direct momentum transfer between liquid and solid instead of a charge-related effect. A more detailed investigation of the role of viscosity is kept for future work.

\paragraph*{Frequency dependency of generated current}
The generated current displays a puzzling dependency on the oscillation frequency : at fixed oscillation amplitude, the current amplitude decreases with frequency making the droplet-graphite system a low-pass filter with a cut-off frequency in the order of 1 Hz (\textit{cf:} Figure S1 \textbf{h}). In agreement with observations under the optical microscope, we attribute this behavior to a thin wetting film spread on the droplet trajectory, which reduces the effective friction (and therefore the generated current). On a 1 s timescale, dewetting takes place and the hemi-spherical droplet reforms.  In this scenario, increasing the frequency is not equivalent to increasing the oscillation amplitude.  \CORR{This dynamic behavior is difficult to probe and rationalize as dewetting timescales strongly depend on surface cleanliness and possible contaminations on the graphene surface. We leave this intriguing dependency for future investigations and as shown on Figure S1 \textbf{h}, report a cut-off frequency in the 1 Hz range.}

\paragraph*{Excluding a modulation effect}
To ensure that the droplet's motion truly generates current and does not merely modulate the DC current driven by the bias voltage, we perform an important control experiment. Instead of using the bias voltage to tune the carrier density, we use a back-gate and measure the electro-fluidic current at zero-bias on single-layer graphene. The results are shown on Figure S1\textbf{f} for a glycerol droplet.

\section*{Appendix D : Wrinkles enhance the electro-fluidic current}

As discussed previously, our sample fabrication techniques allow us to produce graphite flakes of desired thickness with either a high wrinkle density (transferred flakes) or a very flat surface showing negligible roughness (un-transferred flakes). We were therefore able to study the dependency of generated current on both thickness and wrinkle density independently. On Figure S2\textbf{a}, we illustrate the thickness-dependency of the current induced by droplet oscillations of 1 $\mu$m.s$^{-1}$ under a small 20 mV bias voltage for both transferred (diamonds) and un-transferred (circles) graphite flakes. Our experiments reveal that for transferred samples, the thinner the graphite flake, the more efficient the current generation. 

\REF{Like we detailed above and shown on Figures 2\textbf{d} and S2\textbf{a}, the thickness dependency is only apparent and the generated current is in fact determined by the sample's corrugation.} As thinner flakes show a higher corrugation after the transfer, we observe an apparent dependency on flake thickness for the transferred flakes.  Another experimental evidence of the predominant role played by the wrinkles resides in the strong spatial anisotropy of the generated current (\textit{cf:} Figure S2 \textbf{c,d}).  On Figure S2\textbf{c}, we show a cartoon of several oscillation patterns with various angles and mean positions on the graphene flake, the thick arrows show the direction of maximum current at a given position. On \textbf{f}, we show that the generated current depends on both the oscillation angle and mean position (position and angle do not correspond to panel \textbf{c} which is a cartoon).  This variability, linked to the random distribution in height, position and direction of the wrinkles over the graphene strengthens the claim that wrinkles are instrumental to current generation and makes a strong argument in favor of the direct momentum transfer as the source of the current.

\paragraph*{Raman spectroscopy}
To ascertain our claim that wrinkles are folded graphene structures that increase liquid friction and thus motion-induced current by a mere geometrical effect, we perform Raman spectroscopy on one transferred graphite sample and demonstrate that wrinkles show a ratio of D-peak intensity over G-peak intensity of the same order as flatter regions and that they show the usual spectroscopic properties of graphite samples. The measurements are presented on Figure S2 \textbf{g-j},  on \text{g}, we show the AFM scan of a transferred sample. On this sample, the transfer process has torn appart a top layer of graphite which is therefore showing wrinkles on top of the underlying graphite crystal (dark region). We use this particular sample to study the structural properties of wrinkles compared to a flat crystal. On \textbf{h}, we display a typical Raman spectrum measured with a 532 nm laser. On \textbf{i}, we display the G peak intensity map, revealing the in-homogeneity of thickness in the folded regions (higher G peak means thicker region). Finally, on \textbf{d}, we show the ratio of D-peak over G-peak intensity. This shows that wrinkles and folded regions are by no means defects. This analysis strengthen our theoretical treatment of wrinkles and folded regions as graphite or graphene with an enhanced corrugation.

\end{document}